# Thermal Degradation of ligno-cellulosic fuels: DSC and TGA studies

V. Leroy, D. Cancellieri and E. Leoni[*]

SPE-CNRS UMR 6134 University of Corsica

Campus Grossetti B.P 52

20250 Corti (FRANCE).

* : corresponding author

**Mail :** eleoni@univ-corse.fr

**Tel :** +33-495-450-649

**Fax :** +33-495-450-162

**Abstract**

The scope of this work was to show the utility of thermal analysis and calorimetric experiments to study the thermal oxidative degradation of Mediterranean scrubs. We investigated the thermal degradation of four species; DSC and TGA were used under air sweeping to record oxidative reactions in dynamic conditions. Heat released and mass loss are important data to be measured for wildland fires modelling purpose and fire hazard studies on ligno-cellulosic fuels. Around 638 K and 778 K, two dominating and overlapped exothermic peaks were recorded in DSC and individualized using a experimental and numerical separation. This stage allowed obtaining the enthalpy variation of each exothermic phenomenon. As an application, we propose to classify the fuels according to the heat released and the rate constant of each reaction. TGA experiments showed under air two successive mass loss around 638 K and 778 K. Both techniques are useful in order to measure ignitability, combustibility and sustainability of forest fuels.

**Abbreviations:**

$T$: temperature (K)

$\Delta H$: enthalpy of the reaction (endo up) (kJ g$^{-1}$)

$\beta$: heating rate (K min$^{-1}$)

$\Delta m$: variation of mass loss (%)

Ign: ignitability

r: correlation coefficient

F: virgin fuel

[o]: oxidation

G: evolved gases

$k$: rate constant

P: oxidation products

C: chars

A: ashes

$E_a$: activation energy (kJ mol$^{-1}$)

$K_0$: preexponential factor (s$^{-1}$)

$n$: reaction order

$\alpha$: conversion degree

$t$: time (min)

$R$: gas constant 8.314 J mol$^{-1}$

$f(\alpha)$: kinetic model reaction

Comb: Combustibility

**Subscripts:**

melt: melting point

exp: experimental value

Cur: Curie point

p1: peak 1

p2: peak 2

1: refers to exotherm 1

2: refers to exotherm 2

# 1. Introduction

The effect of fire on ecosystems has been a research priority in ecological studies for several years. Nevertheless, in spite of considerable efforts in fire research, our ability to predict the impact of a fire is still limited, and this is partly due to the great variability of fire behaviour in different plant communities [1,2]. Flaming combustion of ligno-cellulosic fuels occurs when the volatile gaseous products from the thermal degradation ignite in the surrounding air. The heat released from combustion causes the ignition of adjacent unburned fuel. Therefore, the analysis of the thermal degradation of ligno-cellulosic fuels is decisive for wildland fire modelling and fuel hazard studies [3-5]. Thermal analysis is widely used in combustion research for both fundamental and practical investigation. Physical fire spread models are based on a detailed description of physical and chemical mechanisms involved in fires. However, thermal degradation models need to be improved. Also, ignitability, combustibility and sustainability of forest fuels are important properties to be determined when talking to efficient wildland fire management. The ignitability determines how easily the fuel ignites, combustibility is the rate of burn after ignition and sustainability counts how well the fuel continues to burn. These properties can be measured by thermal analysis and calorimetric studies of fuels.

Thermal degradation of ligno-cellulosic fuels can be considered according to Figure 1:

**GRAPHIC1.**

There are only a few DSC studies in the literature concerning the thermal decomposition of ligno-cellulosic materials which is preferably followed by TGA [7-10]. It is also important to notice that the literature is very poor in studies concerning the thermal degradation characteristics and kinetics of forest fuels under oxidizing environment. We adapted the DSC in order to measure the heat flow released by natural fuels undergoing thermal decomposition and we used a classical TGA to measure the mass loss of the fuel. Both techniques are useful for the study of thermal degradation since in fire modelling the exchanged energy and the mass loss are fundamental data. Even if the experimental conditions are far from reality, thermal analysis and calorimetric studies can be very helpful in order to compare different fuels behaviour and the combination of those two techniques allows us to improve the knowledge on the thermal degradation of ligno-cellulosic fuels.

We present hereafter the results obtained on different flammable species encountered in the Mediterrannean area.

**2. Experimental**

2.1. Samples preparation

Plant material was collected from a natural mediterranean ecosystem located at 450 meters height above sea level and situated far away from urban areas (near the little town of Corte in Corsica) in order to prevent any pollution on the samples. We chose to study the thermal degradation of rockrose (*Cistus monspeliensis: CM*), heather (*Erica arborea: EA*), strawberry tree (*Arbutus unedo: AU*) and pine (*Pinus pinaster: PP*) which are representative species of the Corsican vegetation concerned by wildland fires. Naturally, the methodology developed hereafter is applicable to every ligno-cellulosic fuel. Aerial parts of each plant were collected in the beginning of April. For each species, a bulk sample from 6 individual plants was collected in order to minimize interspecies differences. Current year, mature leaves were selected, excluding newly developed tissues at the top of the twigs. About 500 g of each species were brought to the laboratory and were sun dried. These materials were stored in Teflon bags and kept at – 18°C. Only small particles (< 5 mm) are considered in fire spread. Also, leaves and twigs were mixed, sampled and oven-dried for 24 hours at 333 K [11]. Dry samples were then grounded and sieved to pass through a 600 µm mesh, then kept to the desiccator. The moisture content coming from self-rehydration was about 4 percent for all the samples. In order to characterize our fuels we decided to perform the elemental analysis on these powders before the thermal analysis. Table 1 shows the results of the elemental analysis performed at the laboratory of the Service Central d'Analyse du CNRS, Lyon, France.

**Table 1**

2.2. Thermogravimetric and calorimetric experiments

We recorded the Heat Flow *vs.* temperature (emitted or absorbed) thanks to a power compensated DSC ( Perkin Elmer®, Pyris® 1) and the mass loss *vs.* temperature thanks to a TGA 6 (Perkin Elmer®).

The DSC calibration was performed out using the melting point reference temperature and enthalpy reference of pure indium and zinc ($T_{melt}$ (In) = 429.8 K, $\Delta H_{melt}$(In) = 28.5 J g$^{-1}$, $T_{melt}$ (Zn) = 692.8 K, $\Delta H_{melt}$(Zn) = 107.5 J g$^{-1}$). Thermal degradation was investigated in the range 473 – 923 K under dry air or nitrogen with a gas flow of 20 mL min$^{-1}$. Samples around 5.0 mg ± 0.1 mg were placed in an open aluminium crucible and an empty crucible was used as a reference. The error caused by weighting gives an error of 1.9 % to 3 % on $\Delta H_{exp}$.

We adapted the DSC for thermal degradation studies by adding an exhaust cover disposed on the measuring cell (degradation gases escape and pressure do not increase in the furnaces). Several experiments were performed with different high heating rates.

The TGA calibration was performed using the Curie point of magnetic standards: perkalloy® and alumel ($T_{Cur}$ (alumel) = 427.4 K, $T_{Cur}$ (perkalloy®) = 669.2 K). Samples around 10.000 mg ± 0.005 mg were placed in an open platinum crucible and the degradation was monitored in the same range of temperature and heating rates as in DSC experiments.

The principal experimental variables which could affect the thermal degradation characteristics in thermal and calorimetric analysis are the pressure, the purge gas flow rate, the heating rate, the weight of the sample and the sample size fraction. In the present study, the operating pressure was kept slightly positive, the purge gas (air or nitrogen) flow rate was maintained at a constant value and the heating rate was varied from β = 10 to 30 K min$^{-1}$. The uniformity of the sample was maintained by spreading it uniformly over the crucible base in all the experiments.

2.3. Thermal separation of DSC records

As it is presented in the results section, the thermal degradation under air of the fuels studied exhibit partially overlapped processes: two exothermic phenomena are recorded in DSC. Thanks to the switching of the surrounding atmosphere in the DSC furnaces we were able to define two independent and successive reactional schemes. The experimental conditions have been modified in order to hide the first exothermic phenomenon. Figure 2 presents the schematic procedure we used to isolate the two phenomena with two experimental steps. The samples were thermally degraded under nitrogen atmosphere (step 1) at different heating rates from 473 K to 923 K. During this step the DSC plots were flats indicating a globally non-exothermic and non-endothermic process. Then the residual charcoal formed during the step 1 was used as a sample to be analyzed by DSC under air sweeping (step 2) with the same temperature range and heating rates as in step 1. Step 1 allowed to pyrolyze the fuels generating a char residue and volatiles which escaped in the surrounding non-oxidizing atmosphere.

**GRAPHIC2**

## 3. Results and discussion

3.1. Thermogravimetric and calorimetric data

Figure 3 shows the experimental DSC/TGA thermograms for an experiment performed at $\beta$ = 30 K min$^{-1}$. In this section, figures present only plots obtained for one heating rate to but two exotherms are clearly visualized and associated with two mass losses for all the heating rates.

**GRAPHIC3.**

During the first exothermic process (around 638 K), gases are emitted and oxidized and the rising temperature contributes to the formation of char. Gases emission are visualized in TGA by a mass loss: 51.3 % < $\Delta m_1$ < 73.2 %. An oxidation of these gases is possible when the surrounding atmosphere selected is air; this phenomenon is represented in DSC by the first exothermic peak. The second exothermic process can be considered like a burning process and it is known as glowing combustion. The char forms ashes in the temperature range of 623 – 823 K, TGA plots show a mass loss 23.1 % < $\Delta m_2$ < 42.8 % and the second exothermic peak is recorded in DSC.

Other authors gave the same ascription for exotherm 1 and exotherm 2 [12,13].

Orfao et al. [14] studied the thermal degradation of pine wood, eucalyptus and pine bark by TGA under air with a linear heating rate of 5 K min$^{-1}$ and they found two successive steps located around 560 K and 685 K. Bilbao et al. [15] recorded two successive mass losses around 590 K and 720 K for pine sawdust studied by TGA with a heating rate of 12 K min$^{-1}$ with different mixtures of Nitrogen/Oxygen as flowing gas. This behaviour has been also reported by Bilbao et al. [16] on cellulose which is the principal component of ligno-cellulosic fuels.

Safi et al. [17] reported two mass loss for the thermal degradation of pine needles under air with a heating rate of 15 K min$^{-1}$. The first one was recorded around 563 – 587 K and the second one was recorded around 701 – 757 K. They also indicated the onset temperature around 521 – 544 K. They also recorded two exothermic peaks in DTA measurements and they ascribed the first (604 – 682 K) to the oxidation of volatiles, while the second peak (739 – 763 K) represents the oxidation of charred residue.

Table 2 presents the DSC records in the range 473 – 923 K, values of enthalpy were obtained by numeric integration on the whole time domain and peak top temperatures were determined thanks to the values of the derivative experimental curve. All the enthalpy values are expressed in kilojoules per gram of the original fuel. The peak top temperatures correspond to the value of temperature at the maximum of the peak. The onset temperatures correspond to the start of oxidation reactions it was automatically determined by the first detected deviation from the baseline curve (tangents plots).

Table 3 presents the results from TGA measurement for the considered heating rates. Mass loss 1 and mass loss 2 refers to the successive thermal events recorded under air. The first mass loss is clearly higher than the second for all the species considered herein. Offset 1 and 2 are the values of temperature corresponding to the end of mass loss 1 and mass loss 2.

**Table 2**

**Table 3**

As expected in classical Thermal Analysis, increasing the heating rate shifts the onset temperature to higher values. However, the results presented in table 2 show that for the heating rates considered in this work, *EA* fuel has the lower onset temperature, followed by *PP*, *CM* and in fine, *AU*. This criterion is very helpful as it can be used as an ignition criterion since onset temperature measure the starting of oxidation reactions. The fuels with low onset temperature are the most ignitable and they burn easily. Our results show that the ignitability is ranked as follows: Ign(*AU*) < Ign(*CM*) < Ign(*PP*) < Ign(*EA*). In DSC, ignition is measured by the onset temperature of the first exothermic peak (i.e. the first detected deviation from the baseline curve).

Table 3 shows that for the heating rates considered in this work, two groups are identified according to the values of mass losses. We named GroupA: *EA* and *PP* fuels and GroupB: *CM* and *AU* fuels. For the GroupA we found $\overline{\Delta m_1} = 72.3\%$ and for the GroupB we found: $\overline{\Delta m_1} = 53.7\%$. TGA measurements can be used in order to classify the fuels according to their capacity to form combustible gases (i.e. ignitability), our results show that GroupA forms more combustible gases than GroupB. Ignitability is directly linked to the quantity of combustible gases emitted by the fuels. So, these results agree with the previous ones (Ign(*AU*) < Ign(*CM*) < Ign(*PP*) < Ign(*EA*)), since the ignitability of fuels from GroupA is greater than the ignitability of fuels from GroupB.

As shown in Table 2, for the heating rates considered in this work, two groups are identified according to the values of enthalpy and peaks top temperature. We named GroupA': *AU* and *EA* fuels and GroupB': *CM* and *PP* fuels. For the GroupA' we found $\overline{\Delta H^o}_{exp} = -12.85 \pm 3\%$ kJ/g, $\overline{T_{p1}} = 641\ K$, $\overline{T_{p2}} = 786\ K$ and for the GroupB', we found: $\overline{\Delta H^o}_{exp} = -10.60 \pm 3\%$ kJ/g, $\overline{T_{p1}} = 633\ K$, $\overline{T_{p2}} = 764\ K$. DSC studies seem to be useful in order to classify the fuels according to their evolved energy when subjected to an external heat flow. Our results show that GroupA' is more energetic than GroupB' so, the sustainability of fuels from GroupA' is greater than the sustainability of fuels from GroupB'.

3.2. Thermal separation of DSC records

The thermal separation of DSC curves in order to isolate each exothermic reaction was performed for all the fuels. During step 1 (see. section 2.3.) we did not observe any thermal effect on the DSC records but TGA plots showed a mass loss during the pyrolysis of the fuels. We attributed this mass loss to the generation of pyrolysis gases.

During step 2, we recorded only one exotherm located in the same temperature range of exotherm 2 which was recorded under air atmosphere. We attributed this exotherm to the oxidation of chars formed during step 1.

Thanks to this thermal separation we can assert that exotherm 1 refers to the oxidation of evolved gases. This oxidation can be recorded as we used a power compensation DSC with micro furnaces and platinum resistance sensors allowing the detection of thermal events in the vicinity of the solid material in the crucible.

Table 4 shows the results of the numerical integration of the second oxidation which where obtained by thermal separation. Enthalpy values for the first oxidation were calculated by substraction from the global enthalpy values.

**Table 4**

3.3. Numerical separation of DSC records

The mathematical interpolation performed with Mathematica® [18] gave equations describing the DSC curves. We fitted the global curves obtained under air with two equations [19], this step have been detailed in a previous work [20].

Thanks to interpolation functions, experimental DSC curves were reconstructed (exotherm 1 and exotherm 2) for all the heating rates considered. Once the exotherms were plotted, enthalpies of each reaction were calculated by numerical integration of the signal and the results are shown in Table 5.

**Table 5**

Since Tab. 4 and Tab. 5 present mean values of three replicate measurements we can conclude that the enthalpy values are constant for each reaction of each plant. We were able to give a mean value for the enthalpy of the gases oxidation (exotherm 1) and for the oxidation of char (exotherm 2) for each species. The obtained values are close whatever the heating rate is.

It is important to notice that the values obtained from the numerical treatment were found to be very close to those obtained by the thermal separation (*cf.* Tab. 4 and Tab 5.).

For every fuels the energy released by the reaction referred to exotherm 2 is more important than the energy released by the reaction referred to exotherm 1. Actually, we found: 4.00 kJ g$^{-1}$ < $|\overline{\Delta H_1^o}|$ < 4.84 kJ g$^{-1}$ for the enthalpy of reaction referred to exotherm 1 whereas we found: 5.93 kJ g$^{-1}$ < $|\overline{\Delta H_2^o}|$ < 8.74 kJ g$^{-1}$ for the enthalpy of reaction referred to exotherm 2.

Figure 4 is an example of experimental data compared to isolated peaks, thanks to the thermal and numerical separation we were able to interpolate with a very good fit the beginning and the end of the global exotherm as it can be seen on figure 4. Figure 5 shows an example of experimental data compared to the interpolated curve which is the sum of peak 1 and peak 2 isolated for each species and each heating rate. For this experiment we obtained a value of r = 0.9946. For all the species investigated and heating rates used the Pearson's correlation coefficient was about this value which indicates a very good fit.

**GRAPHIC 4.**

**GRAPHIC 5.**

Thermal and numerical separation is an indispensable step prior to the kinetic study of each reaction.

3.4. Kinetic study

In our previous work we presented a hybrid kinetic method allowing studying complex and multi-step reactional mechanisms [20]. Here are the results obtained on four species.

The thermal and numerical separation showed that the following kinetic model is suitable:

**GRAPHIC 6**

The first process is modelled as: $F_{(s)} \rightarrow C_{(s)} + G_{(g)}$. The measured heat flow correspond to the oxidation of evolved volatiles (exothermic) in gaseous state ($G_{(g)} \rightarrow P_{1(g)}$). Thus we studied indirectly the kinetics of $F_{(s)} \rightarrow C_{(s)} + G_{(g)}$ by the kinetics of $G_{(g)} \rightarrow P_{1(g)}$ considering $F_{(s)} \rightarrow G_{(g)}$ as the rate limiting reaction of gas production. The second exothermic process concerns the oxidation of chars formed during the first process: $C_{(s)} \rightarrow A_{(s)} + P_{2(g)}$.

Considering two n-th order independent reactions, the kinetic law is expressed as:

$$\frac{d\alpha}{dt} = K_0 \cdot \exp\left(-\frac{E_a}{RT}\right) \cdot (1-\alpha)^n \qquad (1)$$

for each reaction.

We have combined two kind of kinetic methods: model free kinetics and model fitting kinetics.

Model free kinetics is based on an isoconversional method [21-25] where the activation energy is a function of the conversion degree of a chemical reaction. For this work we chose the method of Kissinger-Akahira-Sunose (KAS) applied without any assumption concerning the kinetic model. The KAS method [26] simply consists of extending the Kissinger's method [27] to the conversion range *0.1-0.9*, it is based on Eq. (2):

$$\ln\left(\frac{\beta_i}{T_{jk}^2}\right) = \ln\left(\frac{K_{0\alpha} R}{E_{a\alpha}}\right) - \frac{E_{a\alpha}}{RT_{jk}} - \ln g(\alpha_k) \qquad (2)$$

where $E_{a\alpha}$ and $K_{0\alpha}$ are respectively the apparent activation energy and the pre-exponential factor at a given conversion degree $\alpha_k$, and the temperatures $T_{jk}$ are those which the conversion $\alpha_k$ is reached at a heating rate $\beta_j$. During a series of measurements the heating rate are $\beta = \beta_1...\beta_j...$ The apparent activation energy was obtained from the slope of the linear plot of $\ln\left(\beta_i/T_{jk}^2\right)$ vs. $1/T_{jk}$ performed thanks to a Microsoft® Excel® spreadsheet developed for this purpose.

Model fitting kinetics is based on the fitting of Eq. 1 to the experimental values of *dα/dt*. We used Fork® (CISP Ltd.) software which is provided for model fitting in isothermal or non-isothermal conditions. The resolution of ordinary differential equations was automatically performed by Fork® according to a powerful solver (Runge Kutta order 4 or Livermore Solver of Ordinary Differential Equation). The reaction model $f(\alpha) = (1-\alpha)^n$ was determined among six models specified in the literature [28]. Three heating rates (10, 20, 30 K min$^{-1}$) were used at the same time for each species; the software fit one kinetic triplet and one reaction model valid for all the heating rates. Once the determination of the best kinetic models and optimization of the parameters were achieved, the Residual Sum of Squares between experimental and calculated values indicated the acceptable "goodness of fit" from a statistical point of view.

In order to classify the fuels we chose to use the value of the rate constant at the temperature of the maximum of each exotherm for each fuel. Figures 7 and 8 present the values of rate constants of reaction 1 and reaction 2 calculated at the peak top temperatures with the corresponding mean value of enthalpy reaction. These rate constants were calculated for each species with the best set of kinetic parameters for the three heating rates considered herein.

**GRAPHIC 7**

**GRAPHIC 8**

These results show that the rate constant of gases oxidation is ranked as follow: $k_1(PP) < k_1(AU) < k_1(CM) < k_1(EA)$ and the rate constant of chars oxidation is ranked as follows: $k_2(CM) < k_2(EA) < k_2(AU) < k_2(PP)$

Excepted for *CM* fuel which contain a higher quantity of mineral matter (*cf.* Tab. 1), the limiting step of the thermal degradation is the gases oxidation since $k_1 < k_2$. Then we logically compare the combustibility (the rate of burn after ignition) of the fuels by comparing the constant rate of the gases oxidation, the result is the following: Comb(*PP*) < Comb(*AU*) < Comb(*CM*) < Comb(*EA*).

## 4. Conclusion

Reactions of thermal degradation show multi-step characteristics. We showed that thermal analysis and calorimetric investigations are useful tools in order to get information on the ignitability, combustibility and sustainability of ligno-cellulosic fuels encountered in wildland fires. The data derived from DSC and TGA analysis were: onset temperature and peak top temperature of exothermic peak, global enthalpy of reaction, onset and offset temperatures of successive mass loss. Our results showed that *Erica arborea* and *Pinus pinaster* fuels are the most ignitable species whereas *Arbutus unedo* and *Cistus monspeliensis* are the less ignitable species according to DSC and TGA data. We also found that *Arbutus unedo* and *Erica arborea* are the most energetic species whereas *Pinus pinaster* and *Cistus monspeliensis* are the less energetic species according to DSC data. DSC experiments on such fuels showed two superposed exothermic phenomena. We individualized these phenomena in two oxidative sub-reactions and the enthalpy reaction of each sub-reaction was calculated. We proposed a kinetic scheme for the thermal degradation of ligno-cellulosic fuels under air with nth-order model for the oxidative subreactions observed in DSC. Thanks to an hybrid kinetic method [20] we calculated the rate constant at the peak top temperatures for each species considering two *n-th* order reactions. Our results showed that *Erica arborea* is the most combustible fuel whereas *Pinus pinaster* is the less combustible fuel, *Cistus monspeliensis* and *Arbutus unedo* being intermediate combustible fuels. Even if the experimental conditions of thermal analysis and calorimetry are far from actual conditions of wildland fires we think these tools are useful in order to study the flammability of forest fuels. TGA and DSC experiments allow the determination of combustibility, sustainability and ignitability of fuels and these data should be complimentary to a global flammability parameter obtained thanks to lab-scale flammability test [29,30].

**Acknowledgements**

The authors express their gratitude to the autonomous region of Corsica for sponsoring the present work. This research was also supported by the European Economic Community.

**TABLES**

Table 1: Elemental analysis of the fuels.

|    | C (%) | H (%) | O (%) | Total (%) |
|----|-------|-------|-------|-----------|
| *CM* | 46.58 | 6.22 | 37.68 | 90.48 |
| *EA* | 52.43 | 6.98 | 35.92 | 95.33 |
| *AU* | 48.24 | 6.15 | 40.33 | 94.72 |
| *PP* | 50.64 | 6.76 | 41.53 | 98.93 |

Table 2: Onset temperature, peaks top temperature and global enthalpy measured by DSC.

|  | Species | $\beta = 10$ K min$^{-1}$ | $\beta = 20$ K min$^{-1}$ | $\beta = 30$ K min$^{-1}$ |
|---|---|---|---|---|
| Onset (K) | AU | 563 (1.2) | 567 (1.1) | 572 (1.1) |
|  | EA | 529 (1.4) | 534 (1.8) | 538 (1.5) |
|  | CM | 556 (0.7) | 560 (1.3) | 563 (1.2) |
|  | PP | 540 (2.0) | 546 (0.9) | 549 (1.7) |
| Peak 1 top Temperature (K) | AU | 634 (0.9) | 638 (2.0) | 643 (1.5) |
|  | EA | 639 (1.2) | 644 (1.0) | 649 (1.6) |
|  | CM | 624 (1.3) | 628 (1.6) | 632 (2.3) |
|  | PP | 632 (1.5) | 638 (1.3) | 644 (1.1) |
| Peak 2 top Temperature (K) | AU | 784 (2.0) | 788 (1.2) | 792 (3.0) |
|  | EA | 779 (1.5) | 784 (1.8) | 790 (1.6) |
|  | CM | 772 (1.1) | 776 (2.1) | 780 (1.8) |
|  | PP | 747 (1.1) | 753 (2.2) | 760 (1.3) |
| $\Delta H^o$ (kJ g$^{-1}$) | AU | -13.01 | -12.88 | -12.79 |
|  | EA | -13.02 | -12.96 | -12.46 |
|  | CM | -10.63 | -10.52 | -10.75 |
|  | PP | -10.55 | -10.44 | -10.71 |

Temperature values are the mean values of three replicate measurements and in parenthesis are given the corresponding RSD values

Table 3: Onset temperature, offset temperature and mass loss measured by TGA.

|  | Species | $\beta = 10$ K min$^{-1}$ | $\beta = 20$ K min$^{-1}$ | $\beta = 30$ K min$^{-1}$ |
|---|---|---|---|---|
| Onset (K) | AU | 563 (1.2) | 567 (1.1) | 572 (1.1) |
|  | EA | 529 (1.4) | 534 (1.8) | 538 (1.5) |
|  | CM | 556 (0.7) | 560 (1.3) | 563 (1.2) |
|  | PP | 540 (2.0) | 546 (0.9) | 549 (1.7) |
| Mass loss 1 (%) | AU | 51.3 | 52.7 | 53.4 |
|  | EA | 72.0 | 71.0 | 72.3 |
|  | CM | 54.4 | 54.5 | 56.0 |
|  | PP | 73.2 | 72.2 | 73.0 |
| Offset 1 (K) | AU | 675 (2.3) | 680 (1.0) | 684 (0.6) |
|  | EA | 622 (1.9) | 626 (1.7) | 631 (1.3) |
|  | CM | 638 (1.4) | 642 (1.1) | 648 (1.0) |
|  | PP | 634 (1.3) | 639 (1.3) | 643 (0.5) |
| Mass loss 2 (%) | AU | 42.8 | 41.6 | 42.0 |
|  | EA | 26.0 | 26.6 | 25.7 |
|  | CM | 34.4 | 35.6 | 35.9 |
|  | PP | 23.9 | 23.6 | 23.1 |
| Offset 2 (K) | AU | 803 (1.0) | 807 (1.9) | 812 (1.3) |
|  | EA | 818 (2.0) | 823 (2.1) | 827 (1.9) |
|  | CM | 825 (1.1) | 829 (1.0) | 834 (1.4) |
|  | PP | 809 (1.8) | 813 (1.5) | 818 (1.1) |

Temperature values are the mean values of three replicate measurements and in parenthesis are given the corresponding RSD values

Table 4: Enthalpy values obtained by the thermal separation.

|  | Species | $\beta = 10$ K min$^{-1}$ | $\beta = 20$ K min$^{-1}$ | $\beta = 30$ K min$^{-1}$ | $\overline{\Delta H°}$ |
|---|---|---|---|---|---|
| $\Delta H°_{1\ deducted}$ (kJ g$^{-1}$) | AU | -4.04 | -4.01 | -3.96 | -4.00 (0.04) |
|  | EA | -4.84 | -4.87 | -4.82 | -4.84 (0.07) |
|  | CM | -4.63 | -4.61 | -4.69 | -4.64 (0.05) |
|  | PP | -4.18 | -4.17 | -4.22 | -4.19 (0.03) |
| $\Delta H°_{2\ isolated}$ (kJ g$^{-1}$) | AU | -8.87 | -8.64 | -8.71 | -8.74 (0.10) |
|  | EA | -8.18 | -8.09 | -8.04 | -8.10 (0.08) |
|  | CM | -6.00 | -5.91 | -6.06 | -5.99 (0.08) |
|  | PP | -6.37 | -6.27 | -6.53 | -6.39 (0.14) |

Table 5: Enthalpy values obtained by the numerical separation.

| | Species | β = 10 K min$^{-1}$ | β = 20 K min$^{-1}$ | β = 30 K min$^{-1}$ | $\overline{\Delta H °}$ |
|---|---|---|---|---|---|
| ΔH°$_{1\ calculated}$ (kJ g$^{-1}$) | AU | -4.62 | -4.72 | -4.73 | -4.69 (0.07) |
| | EA | -4.82 | -4.78 | -4.76 | -4.79 (0.03) |
| | CM | -4.62 | -4.55 | -4.64 | -4.60 (0.05) |
| | PP | -4.16 | -4.16 | -4.18 | -4.17 (0.01) |
| ΔH°$_{2\ calculated}$ (kJ g$^{-1}$) | AU | -8.35 | -7.91 | -7.98 | -8.08 (0.27) |
| | EA | -8.16 | -8.00 | -7.99 | -8.05 (0.11) |
| | CM | -5.97 | -5.84 | -5.97 | -5.93 (0.09) |
| | PP | -6.31 | -6.23 | -6.25 | -6.26 (0.05) |

**FIGURES**

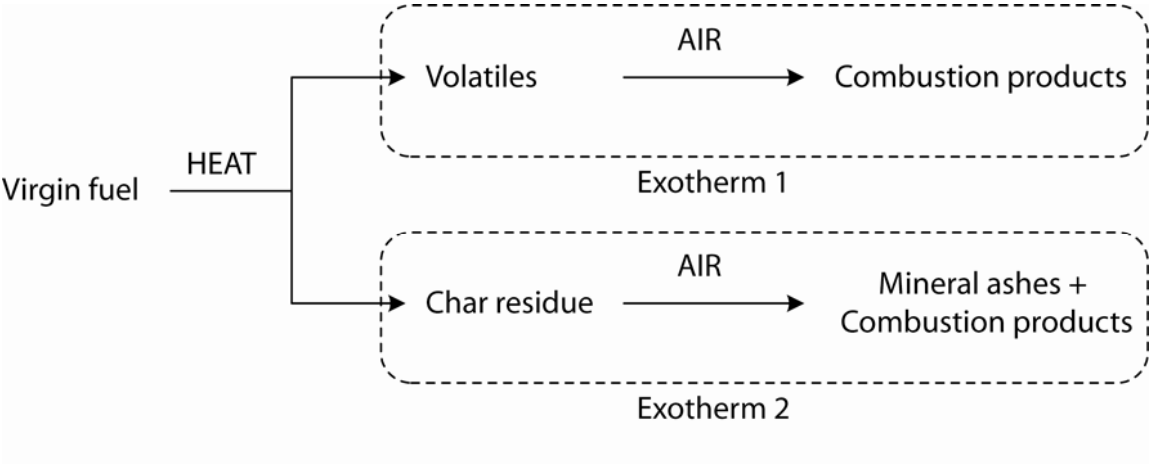

Figure 1: Thermal degradation of a ligno-cellulosic fuel.

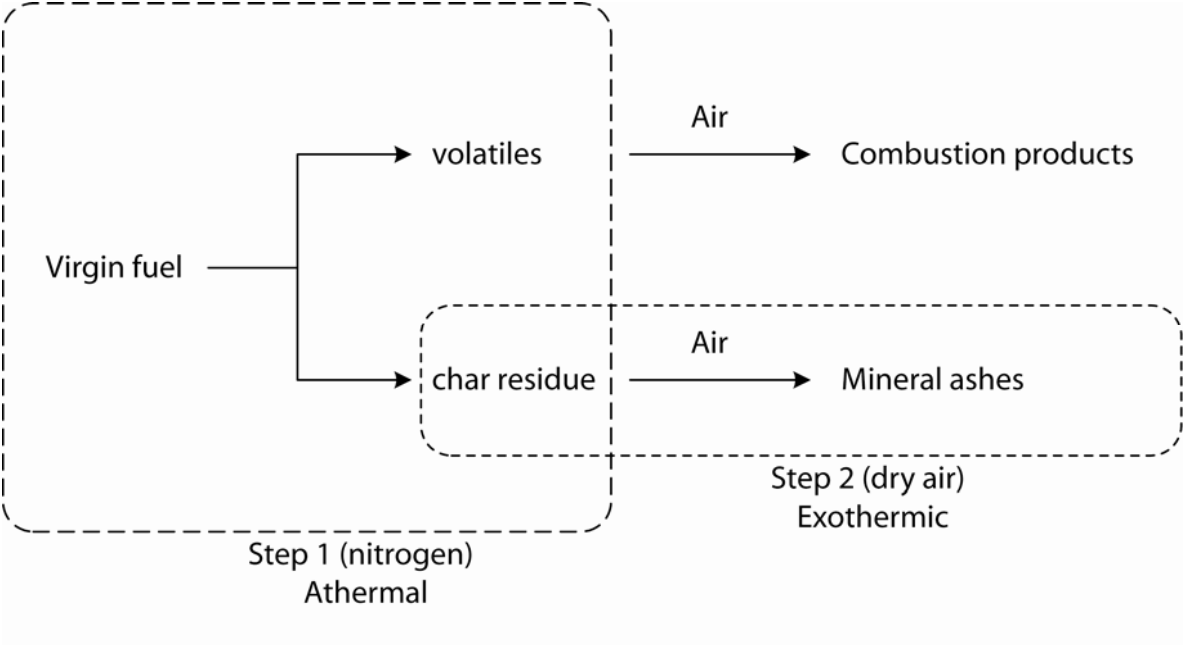

Figure 2: Schematic representation of the thermal separation.

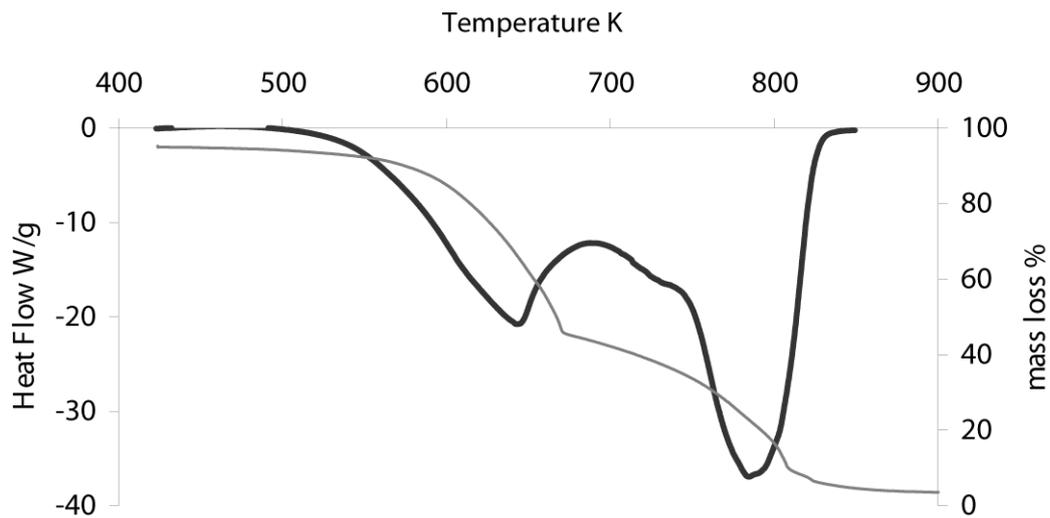

Figure 3: Example of DSC and TGA curves of *EA* fuel obtained with a linear heating rate of 20 K min$^{-1}$ under air atmosphere.

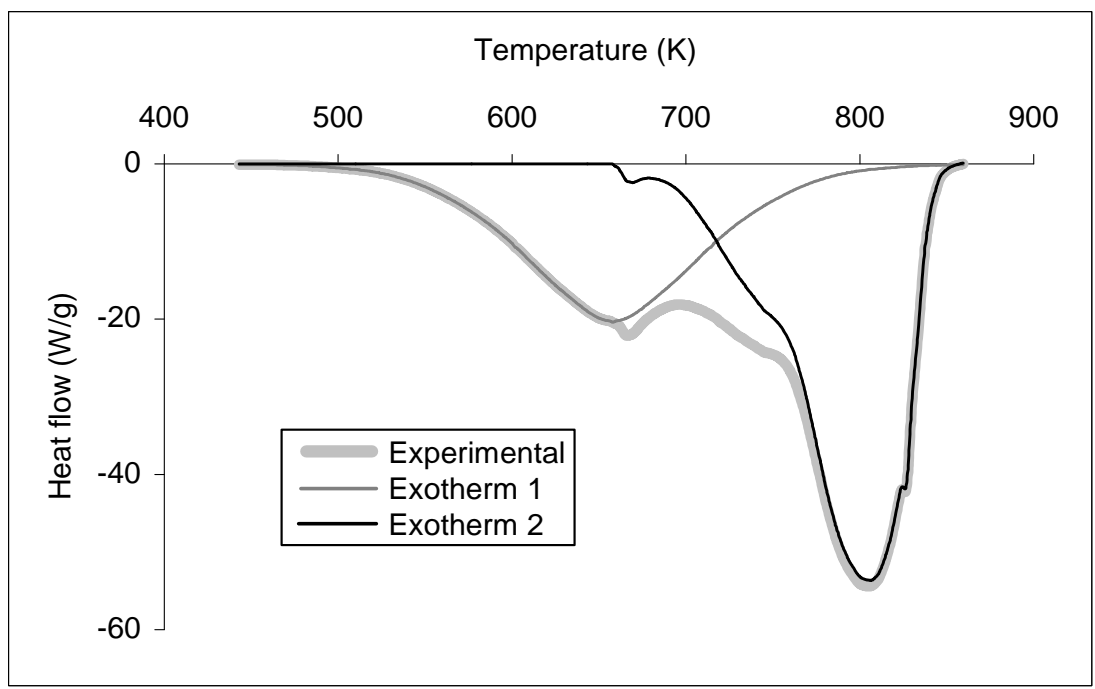

Figure 4: Example of an experimental DSC curve and two isolated peaks (*EA* fuel with a linear heating rate of 30 K min$^{-1}$ under air atmosphere).

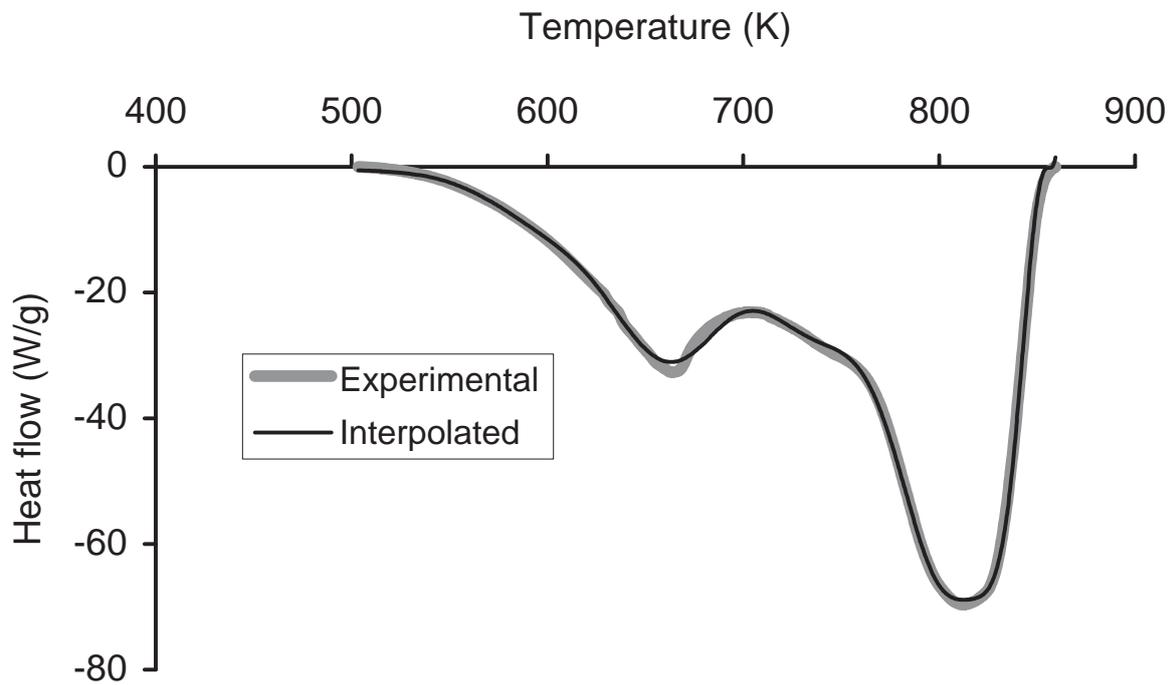

Figure 5: Example of an experimental DSC curve and the interpolated curve (*EA* fuel with a linear heating rate of 20 K min$^{-1}$ under air atmosphere).

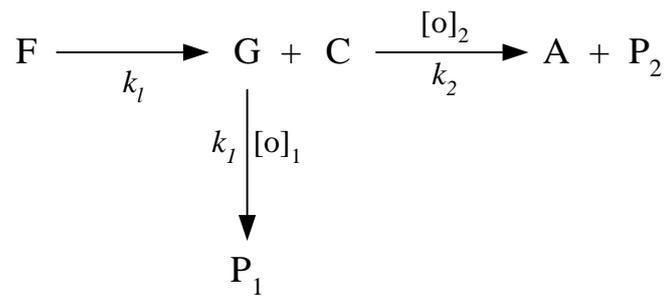

Figure 6: Global kinetic scheme for the thermal degradation of forest fuels.

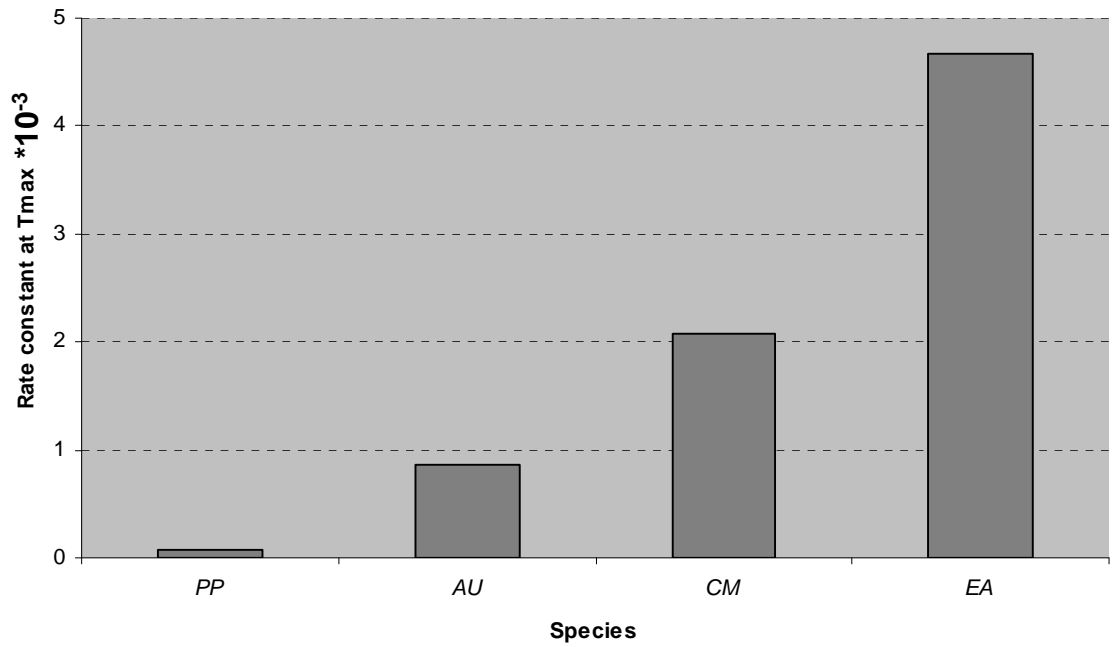

Figure 7: Rate constant for the first step oxidation ($k = K_0 \cdot e^{-\frac{E_a}{RT}}$) calculated at $T_{p1}$ for each species.

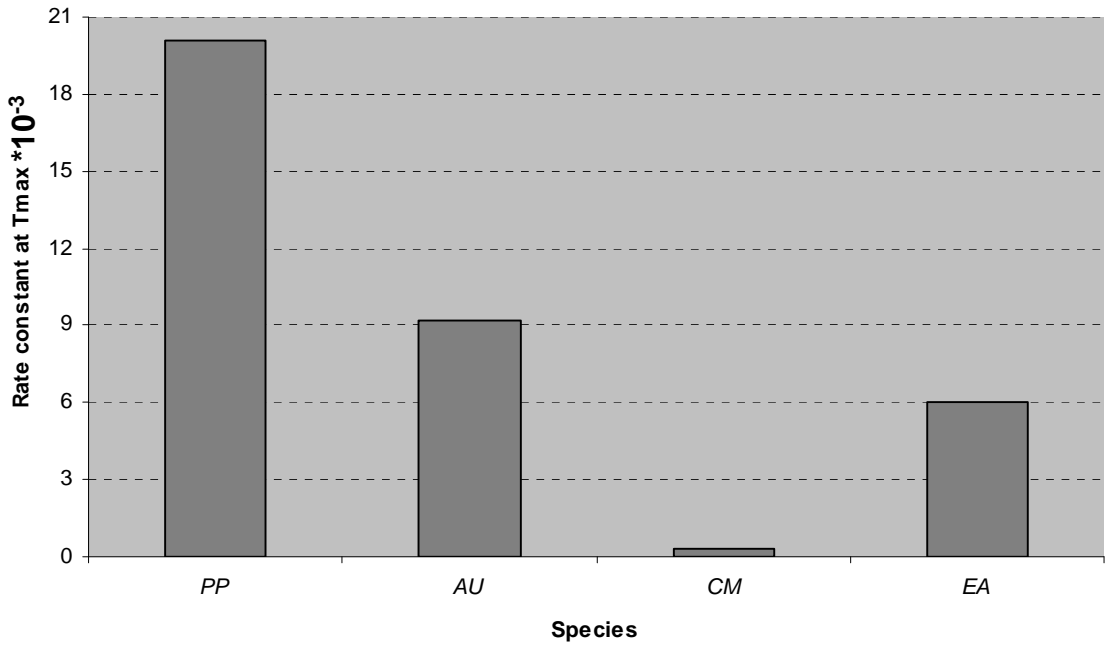

Figure 8: Rate constant for the second step oxidation ($k = K_0 \cdot e^{-\frac{E_a}{RT}}$) calculated at $T = T_{p2}$ for each species.